\documentclass[twocolumn]{revtex4}
\usepackage{graphicx,amsmath,amssymb,amsthm}
\usepackage{color}

\def\identity{\leavevmode\hbox{\small1\kern-3.8pt\normalsize1}}

\renewcommand{\epsilon}{\varepsilon}

\begin{document}

\title{"Commutator formalism" for pairs correlated through Schmidt decomposition as used in Quantum Information}

\author{Monique Combescot}
\affiliation{Institut des NanoSciences de Paris, CNRS, Universit\'e Pierre et Marie Curie, 4 place Jussieu, Paris}
\date{\today}

\begin{abstract}
To easily calculate statistical properties of pairs correlated through Schmidt decomposition, as commonly used in Quantum Information, we propose a "commutator formalism" for these single-index pairs, somewhat simpler than the one we developed for double-index Wannier excitons. We use it here to get the pair number threshold for bosonic behavior of $N$ pairs through the requirement that their number operator mean value must stay close to $N$. While the main term of this mean value is controlled by the second moment of the Schmidt distribution, so that to increase this threshold, we must increase the Schmidt number, higher momenta appearing at higher orders lead to choosing a distribution as flat as possible.
\end{abstract}

\maketitle
Over the last decade, we have extensively studied \cite{CBD} fermion pairs making semiconductor excitons. We have shown that excitons mainly interact through the Pauli exclusion principle between their fermionic components \cite{CB}. This Pauli blocking gives rise to a set of scatterings which correspond to carrier exchanges in the absence of carrier interaction. Being by construction dimensionless, these "Pauli scatterings" control nonlinear effects induced by unabsorbed photons, as easily understood from a bare dimensional argument: scatterings involving Coulomb interaction are energy-like quantities; they must appear with energy denominators which for non-linear optical effects, are of the order of the photon detuning. Detuning of non-absorbed photons being by construction large, these Coulomb processes thus are negligeable in front of pure fermion exchanges. In the same way, just because they are dimensionless, these Pauli scatterings cannot appear in effective hamiltonia
 ns for bosonized excitons \cite{Haug}. As a result, such exciton effective hamiltonians miss a large amount of physical effects, whatever their effective exciton-exciton scatterings.

Usual semiconductor excitons are Wannier excitons. They are made of free electrons and free holes. These double-index excitons $i=(\textbf{Q}_i,\nu_i)$ where $\textbf{Q}_i$ is the center-of-mass momentum and  $\nu_i$ the relative motion index, are linear combination of double-index fermion pairs. Their creation operators read through their wave function in momentum space as
\begin{equation}
B_{i}^{\dagger}=\sum_{\textbf{k}_e,\textbf{k}_h} a_{\textbf{k}_e}^{\dagger} b_{\textbf{k}_h}^{\dagger}\langle\textbf{k}_h,\textbf{k}_e | i \rangle 
 \end{equation}
We have proposed a "commutator formalism" \cite{CBD,CB} to easily handle the consequences of the Pauli exclusion principle between $N$ Wannier excitons. Pauli scatterings  $\lambda(_{mi}^{nj})$ for fermion exchanges between excitons starting in states $(i,j)$ ans ending in states $(m,n)$, in the absence of Coulomb process, formally appear through two commutators, $\left[B_m,B_i^{\dagger}\right]=\delta_{mi}-D_{mi}$ and $\left[D_{mi},B_j^{\dagger}\right]=\sum\Big\{\lambda(_{mi}^{nj})+\lambda(_{ni}^{mj})\Big\}B_n^{\dagger}$. These Pauli scatterings are the keys to explain and better predict physical effects involving unabsorbed photons. They however are rather complex quantities.

 Later on, we turned to Frenkel excitons \cite{CP}. Being made of ion-site excitations, these are single-index excitons. Their creation operators read 
 \begin{equation}
B_{\textbf{Q}}^{\dagger}= \frac{1}{\sqrt{N_s}}\sum_{n} e^{i\textbf{Q}.\textbf{R}_n}a_{n}^{\dagger} b_{n}^{\dagger}
 \end{equation}
 where $N_s$ is the number of ion sites located in a periodic lattice $\textbf{R}_n$. The fact that the Frenkel exciton wave function is just a phase, induces important simplifications on the consequences of Pauli blocking on these correlated pairs.
 
 More recently, we turned to Cooper pairs which also are linear combination of single-index pairs
  \begin{equation}
B^{\dagger}= \sum_{\textbf{k}} a_{\textbf{k}{\uparrow}}^{\dagger} a_{-\textbf{k}{\downarrow}}^{\dagger}(v_k/u_k) 
 \end{equation}
 However, problems raised in BCS superconductivity require not only an exact treatment of the Pauli exclusion principle but also of the potential between up and down spin electrons, to possibly generate the singular potential dependence of the energy \cite{CPC}. This is why the "commutator formalism" we have developped for this problem \cite{CG} stayed at the free pair level.
 
  There is another field in which Pauli blocking between composite bosons plays an important role: Quantum Information. In this field, the correlated pairs are usually written \cite{Law,PL} through their Schmidt  \cite{GRE,LE} decomposition 
 
  \begin{equation}
  B^{\dagger}=\sum_{n} \sqrt{\lambda_n} a_{n}^{\dagger} b_{n}^{\dagger}.
  \end{equation}
  These single-index pairs thus have similarities with Frenkel excitons  although the Schmidt distribution $\sqrt{\lambda_n}$ may not be flat, i.e., it can differ from just a phase. In view of the importance of the field and the difficulty to properly handle the Pauli exclusion principle  through a brut force counting of the amount of blocked events, it is highly desirable to develop a "commutator formalism" appropriate to fermion pairs correlated through a Schmidt decomposition which are the relevant pairs in Quantum Information.
  
 We wish to stress that, according to Schmidt theorem, any two-fermion state can be written through a Schmidt decomposition. The double-index Wannier exciton defined in Eq.(1) would then read $B_i^{\dagger}=\sum_{p} \sqrt{\lambda_{p,i}} \alpha_{p,i}^{\dagger} \beta_{p,i}^{\dagger}$. However, the operators appearing in this decomposition depend on the state $i$ of the exciton at hand. Since, in physically relevant problems dealing with Wannier excitons, the excitons are scattered into different states, the Schmidt decomposition of Wannier exciton creation operators, with different elementary fermion operators for each $i$ exciton, is not appropriate to approach the Wannier exciton physics. This is why we have been led to develop a formalism in which the double-indices of these composite bosons are explicitly kept. Fermion exchanges between excitons appear through Pauli scatterings $\lambda(_{mi}^{nj})$, these dimensionless scatterings allowing Wannier excitons to change states
  from $(i,j)$ to $(m,n)$. The introduction of Pauli scatterings however is unnecessary for single-index pairs. The formalism we here propose is far simpler. It is based on a convenient mathematical quantity, the "generalized correlated pair" defined in Eq.(5). Using it, it is possible to rederive within a few lines, some important Pauli blocking results obtained in the past through far heavier procedures.
  
  In this Letter, we first develop a formalism which allows an easy handling of Pauli blocking between $N$ single-index correlated pairs. We then use it to calculate the mean value of the pair number operator as well as the variance of this number. In a last part, we discuss the appropriate shape of the Schmidt distribution to have a pair number mean value as close to $N$ as possible.

 \textbf{Formalism}:
   We consider free fermion pairs characterized by a single index $l$. Their creation operators read
$B_{l}^{\dagger} = a_{l}^{\dagger} b_{l}^{\dagger}$
where $a_{l}^{\dagger}$ and $b_{l}^{\dagger}$ are creation operators for their fermionic components. The $B_{l}^{\dagger}$'s
are such that $[B_{l'}^{\dagger}, B_{l}^{\dagger}] = 0$ while 
$[B_{l'},B_{l}^{\dagger}] = \delta_{l'l} - D_{l'l}$
where $D_{l'l} = \delta_{l'l}(a_{l}^{\dagger} a_{l} + b_{l}^{\dagger} b_{l})$. 
From them, we construct a set of "generalized correlated pair" operators as 
\begin{equation}
C_{n}^{\dagger} = \sum_{l} |\phi_{l}^2|^{n} \phi_{l} B_{l}^{\dagger}
\end{equation}
with $n = (0, 1, 2, \dots)$ and $\phi_{l}$ normalized by $\sum_{l} |\phi_{l}^2|= 1$. These operators are such that $[C_{m}^{\dagger}, C_{n}^{\dagger}] = 0$, while 
\begin{equation}\label{eqn4}
[C_{m}, C_{n}^{\dagger}] = \tau_{m+n} - D_{m+n}.
\end{equation}
 $\tau_{m}$ is a scalar equal to the 
$(m+1)$-moment
of the $\phi_{l}$ distribution, namely 
\begin{equation}
\tau_{m}=\sum_{l} |\phi_{l}^2|^{m+1}
\end{equation} 
while $D_{m}= \sum_{l}  |\phi_{l}^2|^{m+1}(a_{l}^{\dagger} a_{l} + b_{l}^{\dagger} b_{l})$. 
Since $[a_{l}^{\dagger} a_{l}, B_{l'}^{\dagger}] = \delta_{ll'} B_{l'}^{\dagger} = [b_{l}^{\dagger} b_{l}, B_{l'}^{\dagger}]$, we readily find 
\begin{equation}\label{eqn5}
[D_{m},C_{n}^{\dagger}] = 2 C_{m+n+1}^{\dagger}
\end{equation}

For $N$ correlated pairs with creation operators $C_{0}^{\dagger}$, it is convenient to note that the iteration of Eq.(\ref{eqn5}) leads to
\begin{eqnarray}
[D_{m}, C_{0}^{\dagger N}] = [D_{m}, C_{0}^{\dagger}] C_{0}^{\dagger N-1} + C_{0}^{\dagger} [D_{m}, C_{0}^{\dagger N-1}] \nonumber \\
= 2 N C_{m+1}^{\dagger} C_{0}^{\dagger N-1}\hskip3.1cm
\end{eqnarray}
So, using Eq.(\ref{eqn4}), we get by iteration
\begin{eqnarray}\label{eqn7}
[C_{m}, C_{0}^{\dagger N}] = [C_{m}, C_{0}^{\dagger}] C_{0}^{\dagger N-1} + C_{0}^{\dagger} [C_{m}, C_{0}^{\dagger N-1}] \hskip1.2cm\nonumber \\
= N C_{0}^{\dagger N-1} (\tau_{m} - D_{m}) - N (N-1) C_{0}^{\dagger N-2} C_{m+1}^{\dagger}\hskip0.3cm
\end{eqnarray}
As evidenced below, all statistical properties of $N$ correlated pairs with creation operator $C_{0}^{\dagger}$ follow from this commutator. For the (unnormalized) $N$-pair state 
$|\psi_{N} \rangle = C_{0}^{\dagger N} |0 \rangle$, it in particular gives, since $D_{m} |0 \rangle = 0$
\begin{equation}\label{eqn9}
C_m |\psi_{N} \rangle = N\tau_{m} |\psi_{N-1} \rangle - N (N-1) C_{m+1}^{\dagger} |\psi_{N-2} \rangle.
\end{equation}

\textbf{Number mean value for $N$ correlated pairs}:
Let us write the norm of the $| \psi_{N} \rangle$ state as  
\begin{equation} 
\langle \psi_{N} | \psi_{N} \rangle = \langle 0 |C_0^N C_{0}^{\dagger N}|0 \rangle =N! F_{N} 
\end{equation}
$F_{N}$ is a crucial quantity \cite{CT,CLT} for many-body effects induced by the non-bosonic behavior of composite bosons, "cobosons" in short. $F_N$  would be equal to 1 for $C_0^{\dagger}$ creating an elementary boson. For cobosons, $F_{N}$, still equal to 1 for $N= (0,1)$, decreases when $N$ increases because more and more pair states are excluded from the $C_0^{\dagger}$ sum due to Pauli blocking: this is the so-called "moth-eaten effect". 

We want to determine the mean value of the pair number operator $\hat{N} = C_0^{\dagger} C_0$ in the $ | \psi_{N} \rangle$ state. To do so, it is convenient to first note that, as $C_0^{\dagger} C_{1}^{\dagger} {=} C_{1}^{\dagger} C_0^{\dagger}$ while $\tau_0=1$, Eq.(11) used for $N$ and $N{+}1$ gives \cite{CDD}
\begin{eqnarray}\label{eqn10}
C_0^{\dagger} C_0|\psi_{N} \rangle = N |\psi_{N} \rangle - N (N-1) C_{1}^{\dagger} |\psi_{N-1} \rangle \nonumber \\
= |\psi_{N} \rangle + \frac{N-1}{N+1} C_0 |\psi_{N+1} \rangle.\hskip1.0cm
\end{eqnarray}
From it, we readily get the pair number mean value \cite{CT,Romb} in the $N$-pair state $ |  \psi_{N} \rangle$ as 
\begin{equation}
\langle   \hat{N} \rangle_N{=}\frac{\langle  \psi_{N} | \hat{N} |  \psi_{N} \rangle}{\langle  \psi_{N}  |  \psi_{N} \rangle}
{=}1{+}(N{-}1)\frac{F_{N{+}1}}{F_N}
{=} N (1 {-} \zeta_{N})
\end{equation}
where $\zeta_{N}$ is the fraction of composite bosons which deviates from elementary bosons: The smaller $\zeta_{N}$, the closer to an elementary boson behavior. $\zeta_{N}$ precisely reads
\begin{equation}\label{eqn12}
\zeta_{N} = \frac{N-1}{N} (1 - \frac{F_{N+1}}{F_{N}}).
\end{equation}
This "deviation fraction" reduces to zero for elementary bosons, i.e., $F_{N} = 1$ whatever $N$. It also reduces to zero for $N=1$: two composite particles are needed to evidence their statistical nature. Note that 
$0\leqslant\zeta_{N}\leqslant1$ since, due to the "moth-eaten effect", $F_{N}$ is a decreasing function of $N$. 

Using Eq.(13), we also find
the variance of the particle number as
\begin{eqnarray}
\xi _N{=}\frac{\langle \hat{N}^2 \rangle_N{-}\langle \hat{N} \rangle_N^2}{\langle \hat{N} \rangle_N}\hskip5.3cm
\nonumber \\
{=}\frac{(N{-}1)^2}{N{+}1}\bigg\{ \frac{1{{+}N\frac{F_{N{+}2}}{F_{N{+}1}}-}(N{+}1)\frac{F_{N{+}1}}{F_N}}{1{+}(N{-}1)\frac{F_{N{+}1}}{F_N}}\bigg\}\frac{F_{N{+}1}}{F_N}
\hskip1.cm
\end{eqnarray}
We check that, as $\zeta_N$, this variance reduces to 0 for elementary bosons, $F_N=1$, and also for $N=1$.

When $N$ increases, we physically expect cobosons to "shrink" more and more due to the Pauli exclusion principle between their fermionic components: cobosons become more and more different from a set of single cobosons. As a result, the fraction of cobosons which deviates from an elementary boson behavior should increase with $N$ from $\zeta_{1} = 0$ to $\zeta_{N} \approx 1$ above a certain pair number threshold. In order to determine the $N$ scale over which this occurs, we must explicitly calculate the $F_{N+1}/F_{N}$ ratio which indeed is the key parameter in this problem.  

\textbf{Normalization factor ratio}.
By writing $\langle \psi_{N} |$ as $\langle \psi_{N-1}|C_0$ ,  Eq.(\ref{eqn9}) gives, since $\tau_{0} {=} 1$ 
\begin{equation}\label{eqn13}
\langle \psi_{N} | \psi_{N} \rangle = N \langle \psi_{N-1} | \psi_{N-1} \rangle - N (N-1) \langle \psi_{N-1} | C_{1}^{\dagger} | \psi_{N-2} \rangle.
\end{equation}
 This equation, used for $N+1$, yields
 \begin{equation}
 \frac{F_{N+1}}{F_{N}}=1-N\frac{\langle \psi_{N}| C_{1}^{\dagger}| \psi_{N-1} \rangle}{\langle \psi_{N} | \psi_{N} \rangle}.
 \end{equation}

 This already shows that when the distribution is flat, $\phi_{l} = \exp{(i \varphi_{l})}/\sqrt{N^*}$ for $1 \leq l \leq N^*$, with $N^*$ being the "Schmidt number"and $\phi_{l} =0$ otherwise, 
$C_{1}^{\dagger}$ reduces to $C_0^{\dagger}/N^*$, so that
  \begin{equation}
\Big\{ \frac{F_{N+1}}{F_{N}}\Big\}_{flat}=1-\frac{N}{N^*}.
 \end{equation}
 As a result, 
 $F_{N+1}=0$ for $N\geqslant N^*$:
  the state 
 $ | \psi_{N+1} \rangle$ then reduces to zero. 

 In the case of an arbitrary distribution, we can calculate the $ C_{1}^{\dagger}$ matrix element in Eq.(18) using Eq.(\ref{eqn9}). More generally, this equation gives 
  \begin{eqnarray}
  \langle \psi_{N-1}| C_{m}| \psi_{N} \rangle{=}N\tau_m\langle \psi_{N-1} | \psi_{N-1} \rangle
  \hskip2.5cm
\nonumber \\
{-}N(N{-}1)\langle \psi_{N{-}1}|C_{m+1}^{\dagger} | \psi_{N{-}2} \rangle
\hskip1.cm
   \end{eqnarray}
   Using Eq.(17) for $N\langle \psi_{N-1} | \psi_{N-1} \rangle$,  the above equation allows us to rewrite the $F_N$ ratio as
     \begin{eqnarray}
   \frac{F_{N+1}}{F_{N}}=1-N\tau_1
   \hskip4.5cm
   \nonumber \\
   +N^2(N-1)\frac{\langle \psi_{N-2}| C_{2}-\tau_1C_{1}| \psi_{N-1} \rangle}{\langle \psi_{N} | \psi_{N} \rangle}
     \end{eqnarray}
     For a flat distribution, $C_{m}^{\dagger}=C_{0}^{\dagger}/N^{*m}$ and $\tau_m=1/N^{*m}$; so, the last term in the RHS cancels, in agreement with Eq.(19).
     
     To go further, we iterate the process using Eq.(19). This gives
     \begin{eqnarray}
   \frac{F_{N+1}}{F_{N}}=1-N\tau_1+N(N-1)(\tau_2-\tau_1^2)\frac{F_{N-2}}{F_{N}}
   \hskip1cm
   \nonumber \\
   {-}N^2(N{-}1)^2(N{-}2)\frac{\langle \psi_{N{-}2}| C_{3}^{\dagger}{-}\tau_1C_{2}^{\dagger}| \psi_{N{-}3} \rangle}{\langle \psi_{N} | \psi_{N} \rangle}
    \hskip.5cm
     \end{eqnarray} 
     And so on ... Iteration using Eq.(20) leads to
     \begin{eqnarray}
   \frac{F_{N+1}}{F_{N}}=1-N\tau_1{+}N(N-1)\bigg\{(\tau_2-\tau_1^2)\frac{F_{N-2}}{F_{N}}
   \hskip.5cm
   \nonumber \\
   {-}(N{-}2)(\tau_3{-}\tau_2\tau_1)\frac{F_{N{-}3}}{F_{N}}
    \hskip.5cm
   \nonumber \\
   {+}(N{-}2)(N{-}3)(\tau_4{-}\tau_3\tau_1)\frac{F_{N{-}4}}{F_{N}}{+}\cdots \bigg\}
   \hskip.5cm
        \end{eqnarray}
        
       For a flat distribution, $\tau_n=(1/N^*)^n$; the prefactors of the $F_N$ ratios in the RHS cancel: and we recover Eq.(19). For a general distribution, these prefactors have alternate signs. Indeed, since $\tau_0=1$
       \begin{eqnarray}
       \tau_{m{+}1}{-}\tau_m\tau_1{=}\sum_l |\phi_{l}^{2m{+}4}| \sum_{l'}|\phi_{l'}^2|{-}\sum_l |\phi_{l}^{2m{+}2}| \sum_{l'}|\phi_{l'}^4|
       \hskip.5cm
   \nonumber \\
  =\frac{1}{2}\sum_{ll'}|\phi_{l}^2| |\phi_{l'}^2|\bigg\{|\phi_{l}^2| -|\phi_{l'}^2|\bigg\}\bigg\{|\phi_{l}^{2m}| -|\phi_{l'}^{2m}|\bigg\}\geqslant0
   \hskip.5cm
        \end{eqnarray}
   This shows that the flatter the distribution, the smaller the corrections to the main term $(1-N\tau_1)$.
      
        In view of the $\tau_m$ value for a flat distribution, dimensional arguments lead to  $\tau_m$ scaling as $(1/N_{eff}^*)^m$ where $N_{eff}^* =1/\tau_1$  is the "effective Schmidt number" of the distribution at hand. For $N$ small compared to this number, Eq.(23) gives the $F_N$ ratio as \cite{Ravi}
        \begin{equation}
         \frac{F_{N+1}}{F_{N}}\simeq1-N\tau_1+N^2(\tau_2-\tau_1^2)-N^3(\tau_3-3\tau_2\tau_1+2\tau_1^3)-\cdots
         \end{equation}

         For arbitrary $N$, it is possible to show that the sum of terms in $N(N-1)$ in Eq.(23) gives a positive contribution to the $F_N$ ratio, so that this ratio is larger than $(1-N\tau_1)$
  in agreement with the beautiful inequality recently derived by Wootters's group \cite{Wot}, namely 
 \begin{equation}
        1-N\tau_1\leqslant \frac{F_{N+1}}{F_{N}}\leqslant1-\tau_1
         \end{equation}
To show it, we come back to Eq.(21). By noting that the $C_1^{\dagger}$ matrix element in Eq.(17) must be real, we can rewrite the $N(N-1)$ factor in Eq.(21) as
\begin{equation}
N\langle \psi_{N-2}| C_2|\psi_{N-1} \rangle-N\tau_1\langle \psi_{N-1}|C_{1}^{\dagger}| \psi_{N-2} \rangle
  \end{equation}
  We then replace $N\tau_1\langle \psi_{N-1}|$ according to Eq.(9). This leads to
 \begin{eqnarray}
 \langle \psi_{N-2}|C_2\bigg\{N|\psi_{N-1} \rangle-N(N-1)C_{1}^{\dagger} \psi_{N-2} \rangle\bigg\}
    \hskip.5cm
   \nonumber \\
   -\langle \psi_{N}| C_{1}^{\dagger2}| \psi_{N-2} \rangle
   \hskip2.5cm
  \end{eqnarray}
  in which we again use Eq.(11) to replace the bracket by  $ C_0| \psi_{N} \rangle$. As a result, Eq.(21) also reads  
   \begin{eqnarray}
   \frac{F_{N+1}}{F_{N}}=1-N\tau_1
   \hskip4.5cm
   \nonumber \\
   +N(N-1)\frac{\langle \psi_{N-1}| C_2|\psi_{N} \rangle-\langle \psi_{N}|C_{1}^{\dagger2}| \psi_{N-2} \rangle}{\langle \psi_{N} | \psi_{N} \rangle}
     \end{eqnarray}
   The simplest way to show that the last term is positive is to expand it on free pair operators. We then find that the $\phi_{l}$ distribution appears through
  \begin{eqnarray}
    \sum_{l_1\cdots l_N}^{(\neq)} |\phi_{l_1}^{2}|^{3}  |\phi_{l_2}^{2}| \cdots  |\phi_{l_N}^{2}|{-}\sum_{l_1\cdots l_N}^{(\neq)} |\phi_{l_1}^{2}|^{2}  |\phi_{l_2}^{2}|^ {2}|\phi_{l_3}^{2}| \cdots  |\phi_{l_N}^{2}|
           \hskip.0cm
   \nonumber \\
  =\frac{1}{2}\sum_{l_1\cdots l_N}^{(\neq)}|\phi_{l_1}^2|  \cdots  |\phi_{l_N}^{2}|\bigg\{|\phi_{l_1}^2| {-}|\phi_{l_2}^2|\bigg\}^2\geqslant0
   \hskip0.8cm
        \end{eqnarray}
        the sums being taken over differents $(l_1\cdots l_N)$. As a result, $F_ {N+1}/F_N$ is larger or equal to $1-N\tau_1$.
        
        To derive the upper bound in the inequality (26), we can also use Eq.(29). We then have to show that
         \begin{eqnarray} 
(N-1)\tau_1\langle \psi_{N} | \psi_{N} \rangle
         \hskip4.5cm
   \nonumber \\
  +N(N-1)\bigg\{ \langle \psi_{N}|C_{1}^{\dagger2}| \psi_{N-2} \rangle-\langle \psi_{N-1}| C_2|\psi_{N} \rangle\bigg\}  
 \end{eqnarray}
 is positive. Using Eq.(20) for $\tau_1\langle \psi_{N} | \psi_{N} \rangle$ and noting that the $C_2$ matrix element is real as seen from Eqs.(18,21), the above quantity also reads
  \begin{equation} 
  \frac{N-1}{N+1}\langle \psi_{N}| C_1 | \psi_{N+1} \rangle+N(N-1)\langle \psi_{N}|C_{1}^{\dagger2}| \psi_{N-2} \rangle
  \end{equation} 
  which is the sum of two positive terms as easily seen by expending them on free pair operators.
We wish to note that the upper bound in the inequality (26) just corresponds to $F_ {N+1}/F_N$ being a decreasing function of $N$ since $F_2/F_1=(1-\tau_1)$ as seen from Eq.(21). Actually, this $N$ independent upper bound leads to a "deviation fraction" $\zeta_N$ larger than $\tau_1(1-1/N)$  which is not really useful to determine the $N$ threshold above which cobosons substantially deviate from bosonic behavior.

 Although the above formalism provides a very direct way to reach the $F_{N+1}/F_N$ ratio relevant in the problem we here consider, we wish to mention that it is possible to recover this ratio through the general equation which links the various $F_N$'s. A simple way to get this equation which does not use the Pauli scatterings $\lambda(_{mi}^{nj})$ introduced in previous derivations, is to start with Eq.(17) and use Eq.(20). We then get
\begin{eqnarray}
\langle \psi_{N} | \psi_{N} \rangle = N \langle \psi_{N-1}|\psi_{N-1} \rangle
- N (N-1)^2 \tau_{1} \langle \psi_{N-2} | \psi_{N-2} \rangle \nonumber \\\hskip-3.3cm
+ N (N-1)^{2} (N-2)^{2} \langle \psi_{N-3} | C_{2} | \psi_{N-2} \rangle.\hskip0.3cm 
\end{eqnarray}
Iteration leads to an equation between the $F_{N}$'s only which reads as 
\begin{equation}\label{eqn15}
F_{N} = F_{N{-}1} {-} (N{-}1) \tau_{1} F_{N{-}2}{+} (N{-}1) (N{-}2) \tau_{2} F_{N{-}3}{-}\dots%
\end{equation}
This gives the first $F_{N}$'s as $F_{1} {=} \tau_{0}{=} 1$, $F_{2} {=} 1 {-} \tau_{1}$, $F_{3} {=} 1 {-} 3 \tau_{1} {+}2 \tau_{2}$, 
the general expression of $F_N$ reading a
\begin{equation}\label{eqn17}
F_{N} =1{-}N(N{-}1)\bigg\{ \mu_{1}{-}(N{-}2)\mu_{2}
{+}(N{-}2)(N{-}3)\mu_{3}\cdots\bigg\}
\end{equation}
with $\mu_{1}=\tau_{1}/2$, $\mu_{2}=\tau_{2}/3$, $\mu_{3}=\tau_{3}/4-\tau_{1}^2/8$, and so on... as can be checked by inserting Eq.(35) into Eq.(34).
This shows that $F_N$ has an overextensive dependence on $N$, with terms in $Nf(N/N_{eff}^*)$, this overextensive dependence however disappearing in physical quantities which depends on $F_N$ ratios only.

\textbf{"Deviation fraction"}
We now come back to the "deviation fraction" defined in Eq.(15) and consider a few particular distributions.

(i)\textit{ Flat distribution}:
Using the value of the $F_{N+1}/F_N$ ratio for a flat distribution given in Eq.(19), we readily find that the "deviation fraction" then takes a compact form, namely 
\begin{equation}\label{eqn19}
\big\{\zeta_{N}\big\}_{flat}= (N-1)/N^*.
\end{equation}
We see that this "deviation fraction" $\zeta_{N}$ increases when $N$ increases from its bosonic value $\zeta_{1}=0$ to $\zeta_{N^*}=(1 - 1/N^*)$  and then stays equal to 1 for $N \geq(N^* + 1)$, all cobosons differing from elementary bosons above this threshold. Eq.(36) shows that,
for a flat distribution, the larger the Schmidt
number $N^*$, the larger the number of cobosons which possibly behave as elementary bosons.  If we now turn to the variance of the particle number defined in Eq. (16), we find that, for a flat distribution, it reduces to zero for all $N$, as in the case of elementary bosons
\begin{equation}\label{eqn19}
\big\{\xi_{N}\big\}_{flat}= 0.
\end{equation}

\textit{(ii) Peak or canyon distribution}:
We now consider a distribution with the same Schmidt number $N^*$ , i.e., $|\phi_{l}| \neq 0$ for $1\leqslant l \leqslant N^*$ but with $|\phi_{l}|$ now peaked or depressed on $l = l_{c}$. It is clear that in this case too, Pauli blocking imposes $C_0^{\dagger N}|0\rangle=0$ for $N > N^*$. The question is to know if, for a given $N^*$, a peak distribution would help to decrease the fraction of $N$ cobosons which deviates from bosons. Intuitively, the answer should be no, because a peak distribution can be approximated by a step function which fundamentally corresponds to a reduction  of the Schmidt number.

To illustrate it, let us consider a very simple case: all the $|\phi_{l}|^{2}$'s are equal except one, namely $|\phi_{l }|^{2} = (1-x)/N^{*}$ for all $l \neq l_{c}$ in the $(1,N^*)$ range. $\tau_{0} = 1$ then imposes $|\phi_{l_{c}}|^{2} = x+(1-x)/N^{*}$; so, positive $x$'s correspond to a peak and negative $x$'s to a canyon. The fact that $|\phi_{l \neq l_{c}}|^{2}$ and $|\phi_{l_{c}}|^{2}$ must be both positive restricts $x$ to $-1/(N^{*}-1) < x < 1$. The $x$ upper bound corresponds to only having the $l_{c}$ state populated while the $x$ lower bound corresponds to having this $l_{c}$ state empty. The second and third momenta of this distribution are given by 
\begin{eqnarray}\label{eqn20}
N^{*} \tau_{1} = 1 + x^{2} (N^{*} - 1)
\hskip3.8cm
 \nonumber \\
N^{*2} \tau_{2} = 1 - 3 x^{2} (N^{*} - 1) + x^{3} (N^{*} - 1) (N^{*} - 2).
\end{eqnarray}

Eqs.(15,38) then give the "deviation fraction" for two correlated pairs as 
\begin{eqnarray}
\zeta_{2} = \frac{\tau_{1} - \tau_{2}}{1 - \tau_{1}} = \frac{1}{N^{*}} \bigg[1 + (N^{*} - 2)\frac{x^2}{1+x}\bigg]. 
\end{eqnarray}
We recover the flat distribution result of Eq.(36) for $x$ = 0, the above equation then giving $\zeta_{2} = 1/N^{*}$, and also for $x = -1/(N^{*}-1)$ which corresponds to $|\phi_{l_{c}}|^{2} = 0$: we then only have $(N^{*} - 1)$ states in the entanglement; so, $\zeta_{2} = 1/(N^{*} - 1)$. Eq.(39) shows that, in the range of physical $x$'s, the fraction of two cobosons which deviates from elementary bosons is minimum for $x = 0$, i.e., when the $|\phi_{l}|^{2}$ distribution is flat as physically guessed: to deviate from a flat distribution, by either making a narrow hole ($x<0$) or a narrow peak ($x>0$) tends to increase this fraction. 

\textit{(iii) Arbitrary distribution}:
We now turn to an arbitrary distribution. The $\tau_{n}$ momenta decrease from $\tau_{0} = 1$ when $n$ increases. The broader the $|\phi_{l}|^{2}$ distribution, the faster the decrease, $\tau_{n}$ scaling as $(1/N_{eff}^*)^{n}$ where $N_{eff}^*$ is the "effective Schmidt number". 
To possibly have a large number of cobosons behaving as elementary bosons, the distribution must be broad enough for $\tau_{n}$ to be much smaller than 1 for $n=1$ already. For $N\gg1$ but still small enough for $N/N_{eff}^*$ to be small, Eq.(25) gives the "deviation fraction" as 
\begin{equation}
\zeta_{N} \approx N \tau_{1} - N^{2} (\tau_{2} - \tau_{1}^{2}) + N^{3}(\tau_{3} -3\tau_{2}\tau_{1}+2\tau_{1}^3)- \dots
\end{equation}
We see that, since $\tau_{n} \propto (1/N_{eff}^*)^{n}$, the first term of $\zeta_{N}$ scales as $N/N_{eff}^*$, the second term as $(N/N_{eff}^*)^{2}$ and so on... 
This shows that the scale for the number of cobosons possibly behaving as elementary bosons is essentially related to the effective Schmidt number through the second moment $\tau_{1}$ of the Schmidt distribution, often called purity: the larger $N_{eff}^*=1/\tau_{1}$, the larger the correlated pair number for which $\zeta_{N} \ll 1$. This physically leads to take the largest possible Schmidt number, corrections to the first term of Eq.(40), which depend on higher momenta, leading us to choose a $\phi_l $ distributin as flat as possible. 

\textbf{To conclude}, we have constructed a "commutator formalism" for pairs correlated through the Schmidt decomposition which are the relevant pairs in Quantum Information. Through its key equation (11), this formalism allows us to recover results linked to Pauli blocking obtained previously in various different contexts ( Wannier excitons \cite{CBD}, Frenkel excitons \cite{CP}, Cooper pairs \cite{CG} and also Quantum Optics \cite{CDD}) through far heavier procedures. 
This formalism is definitly valuable for the reader not to be forced to go through more complicated derivations which only are of interest for problems dealing with double-index correlated pairs.
 The present formalism makes use of a "generalized correlated pair" creation operators $C_{n}^{\dagger}$ which are convenient mathematical quantities for easy calculations envolving $N$ correlated pairs with same creation operator $C_{0}^{\dagger}$.
 
 We first use this "commutator formalism" to calculate the ratio of normalization factors $F_{N+1}/F_N$ in terms of the various momenta of the Schmidt distribution, this ratio being the relevant quantity for physical effects induced by the Pauli exclusion principle between $N$ identical  correlated pairs. This formalism allows us to also rederive the beautiful inequality for the 
 $F_{N+1}/F_{N}$ ratio recently obtained by Wooters's group \cite{Wot} through a careful counting of the amount of terms left after Pauli blocking in combination of scalar products like $\langle \psi_{N} | \psi_{N} \rangle$'s. 
 
 We finally use this formalism to determine the Scmidt distribution for which $N$ cobosons would be close to $N$ elementary bosons, by looking at the mean value of their number operator. We find that for a flat distribution, i.e., for a distribution which is just a phase factor, the fraction of $N$ cobosons which deviates from an elementary boson behavior takes a very compact form, in terms of the second moment of the Schmidt distribution, the variance of this particle number then reducing to zero. In the general case, this "deviation fraction" has a more complex form in which  enter higher momenta of the Schmidt distribution. The study of a "peak" or "canyon" distribution shows in a transparent way that, while an increase of the Schmidt number always is favorable, the flatter the distribution, the larger the amount of cobosons behaving as elementary bosons.

{\it Acknowledgement.}
I wish to thank Vlatko Vedral and Shen Yong Ho for inviting me to give a set of lectures on the many-body theory of composite bosons, at the Center for Quantum Technologies of the National University of Singapore. I also wish to thank Dagomir Kaszlikowski and Ravishankar Ramanathan for introducing me to the field of Quantum Information. I have benefited from valuable discussions on the importance of the composite boson aspect of entangled pairs, along a preprint I have seen prior its publication (Ref. 15). This gave me the idea to, as done in the present work, adapt to single-index pairs correlated through a Schmidt ditribution, the coboson formalism first developed fo double-index excitons.

\bibliographystyle{naturemag}

\end{document}